# Enhanced Thermal Object Imaging by Photon Addition or Subtraction

Claudio G Parazzoli[1], Benjamin E. Koltenbah[1], David R. Gerwe[1], Paul S. Idell[1], Bryan T. Gard[2], Richard Birrittella[3], S. M. Hashemi Rafsanjani[4], Mohammad Mirhosseini[4], O. S. Magana-Loaiza[4], Jonathan P. Dowling[2], Christopher C. Gerry[3], Robert W. Boyd[4], Barbara A Capron[1]

[1]The Boeing Company, P.O. Box 3707 Seattle, WA. 98124, [2]Hearne Institute for Theoretical Physics and Department of Physics & Astronomy Louisiana State University, Baton Rouge LA 70803, [3]Lehman College, The City University of New York, [4]The University of Rochester, Rochester, NY 14627

**Abstract:** Long-baseline interferometry (LBI) is used to reconstruct the image of faint thermal objects. The image quality, for a given exposure time, is in general limited by a low signal-to-noise ratio (SNR). We show theoretically that a significant increase of the SNR, in a LBI, is possible by adding or subtracting photons to the thermal beam. At low photon counts, photon addition-subtraction technology strongly enhances the image quality. We have experimentally realized a nondeterministic physical protocol for photon subtraction. Our theoretical predictions are supported by experimental results.

Imaging of faint objects, illuminated by thermal light from natural sources, is a challenging problem in scientific, civilian, and military arenas. The problem becomes more acute for short exposure times, or equivalently, low average photon occupation number. The principles of image reconstruction were first suggested by Michelson[1] and later codified in the van Cittert-Zernike theorem[2]. The theorem states that the source intensity distribution, $I(x,y)$, is the Fourier transform of the equal time complex degree of coherence $j(u,v)$.

$$I(x,y) = \iint_{u,v} j(u,v) e^{ik[(x-u)+(y-v)]} du\, dv \Big/ \iint_{u,v} j(u,v)\, du\, dv \quad (1)$$

In a long-baseline interferometer (LBI) the thermal photons from a distant source are collected by two apertures, A1 and A2. A1 is at a fixed position $P(0,0)$ in the u-v plane, and A2 is at $P(u,v)$. As A2 scans the u-v plane, the difference of the accumulated phase $\phi$ along the paths from the source to A1 and A2, is modulated.

When the thermal photons from A1 and A2 are interfered on a beam splitter, the modulation of $\phi$ generates the fringe pattern $j(u,v)$. The SNR of the fringe pattern $j(u,v)$, $SNR_j$, is related to $SNR_{Th}$ of the average occupation number, $\bar{n}$.

$SNR_{Th} = \bar{n}/\sqrt{\bar{n}(1+\bar{n})}$, and for $\bar{n} \ll 1$, $SNR_{Th} \triangleright \sqrt{\bar{n}}$. At low $\bar{n}$ $SNR_j$ decreases, and to obtain high resolution images an increasing exposure time is required. In this paper we address the problem of increasing $SNR_j$ above the thermal limit of $SNR_j @ \sqrt{\bar{n}}$ and thus reducing the exposure times for given image resolution.

The problem of increasing $SNR_j$ above its thermal limit is closely related to minimization of the standard deviation $\sigma_\phi$ of the phase difference $\phi$ between the two arms of a Mach-Zehnder interferometer. A good summary of recent work on this subject is found in Ref. [3].

We studied the effects of photon addition or subtraction technology (PAST) on one of the output ports of the LBI beam splitter where the fringes are generated, as shown in Figure 1. The PAST idea was first suggested, in a different context, in Ref. [4]. To implement the PAST concept, the state density matrix $\hat{\rho}$ is formally modified by the addition or subtraction process as[5]

$$\hat{\rho}^+ = (\hat{a}^\dagger)^m \cdot \hat{\rho} \cdot (\hat{a})^m / N^+, \quad N^+ = \mathrm{Tr}[(\hat{a}^\dagger)^m \cdot \hat{\rho} \cdot (\hat{a})^m], \quad m \geq 0$$
$$\hat{\rho}^- = (\hat{a})^{-m} \cdot \hat{\rho} \cdot (\hat{a}^\dagger)^{-m} / N^-, \quad N^- = \mathrm{Tr}[(\hat{a})^{-m} \cdot \hat{\rho} \cdot (\hat{a}^\dagger)^{-m}], \quad m \leq 0$$
$$(2)$$

We determined theoretically that PAST significantly improves $SNR_j$ and consequently the resolution of the image. We refer to the formulation of PAST, given in Eq. (2), as the mathematical protocol (MP).

In the theoretical approach, we considered only the single-mode case. As we discuss below, this is not a restrictive assumption for $\bar{n} \cong 1$. The single-mode thermal probability distribution of photons, a Bose-Einstein (B-E) distribution, is $p(n) = \bar{n}^n/(1+\bar{n})^{n+1}$

For a rectangular spectral distribution, all the modes share the same average occupation number $\bar{n}$, the distribution is [2],

$$p(n,\mu,\bar{n}) = \frac{(n+\mu-1)!}{(\mu-1)!n!}(1+\bar{n}/\mu)^{-\mu}(1+\mu/\bar{n})^{-n}$$

It can be shown that $p(n,\mu,\bar{n})$ approaches the B-E distribution for $\bar{n} \cong 1$ and arbitrary $\mu$.

This behavior justifies using the single mode case in our study where $\bar{n} \cong 1$.

The LBI simulation geometry is schematized in Fig. 1. It is equivalent to a Mach-Zehnder Interferometer (MZI).

A single-mode thermal beam of infinite time duration, with an average occupation number of $2\bar{n}$, enters one port of a 50-50 beam splitter and the other port is open to vacuum. The phase shift $\phi$ in the upper arm of the MZI, represents the phase accumulation difference between the paths from the object to A1 and A2. The entities within the yellow box in Figure 1 emulate the thermal object whose output is collected by two separated apertures in the LBI. The input density matrix $\hat{\rho}_{In} = \hat{\rho}_{Th} \otimes \hat{\rho}_{Vac}$ has been evaluated in terms of the MZI output states $|n_5\rangle$ and $|n_6\rangle$ using standard methods, and with the help of a quantum simulation package in Mathematica[6]. The resulting density matrix at the output of the second balanced beam splitter BS2 is

$$\hat{\rho}_{Out} = \sum_{n=0}^{\infty} \frac{(2\bar{n})^n}{(1+2\bar{n})^{n+1}} |n_{\hat{1}}, 0_{\hat{2}}\rangle \cdot \langle n_{\hat{1}}, 0_{\hat{2}}|$$

$$= \sum_{n=0}^{\infty} \sum_{k=0}^{n} \sum_{k'=0}^{n} \frac{(2\bar{n})^n}{(1+2\bar{n})^{n+1}} \frac{n!}{\sqrt{(n-k)!\,(n-k')!\,k!\,k'!}}$$

$$\times (c_5)^{n-k} (c_5^*)^{n-k'} (c_6)^k (c_6^*)^{k'} |(n-k)_{\hat{5}}, k_{\hat{6}}\rangle \cdot \langle (n-k')_5, k'_6|$$

with $c_5 = -(1-e^{i\phi})/2$ and $c_6 = i(1+e^{i\phi})/2$. The subscripts refer to modes 5 and 6 in Fig. 3. Next we apply the MP described by Eq. (2) and set $\hat{\rho}_m^+$ and $\hat{\rho}_m^-$ to be the density matrices corresponding to the addition or subtraction of $m$ photons respectively, $m \geq 0$ for addition, $m \leq 0$ for subtraction.

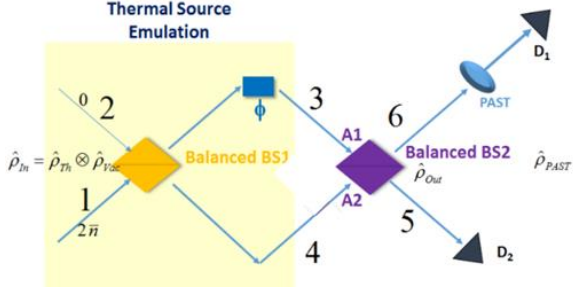

FIG. 1 Schematic used in the computation to emulate the LBI

By tracing over mode 5 we obtain the photon distributions at detector D1, $P_6^+$ and $P_6^-$ for photon addition and subtraction respectively.

$$P_6^+(n_6) = \sum_{n_5=0}^{\infty} P_{56}^+(n_5, n_6)$$

$$= \frac{n_6!}{m!(n_6-m)!} \frac{(2\bar{n}\cos^2(\phi/2))^{n_6-m}}{(1+2\bar{n}\cos^2(\phi/2))^{1+n_6}}$$

$$P_6^-(n_6) = \sum_{n_5=0}^{\infty} P_{56}^-(n_5, n_6),$$

$$= \frac{(n_6-m)!}{(-m)!\,n_6!} \frac{(2\bar{n}\cos^2(\phi/2))^{n_6}}{(1+2\bar{n}\cos^2(\phi/2))^{1+n_6-m}}$$

(7)

The average occupation numbers at D1 for photon addition or subtraction are

$$\langle N_6 \rangle_m^+ = 2\bar{n}\cos^2(\phi/2) + m(1+2\bar{n}\cos^2(\phi/2)), \quad m \geq 0$$
(3a)

$$\langle N_6 \rangle_m^- = (1-m)2\bar{n}\cos^2(\phi/2), \quad m \leq 0$$
(3b)

Two interesting results emerge from Eqs. (3). First, from Eq. (3a) for $\phi = 0$, all the photons are collected at D1. This is the standard result for a balanced MZI. Secondly, either for photon addition or subtraction, the average occupation number is augmented. The increase in the average occupation number is larger for photon addition than subtraction. This is the result of the modified photon distribution that happens by adding or subtracting photons. The modification of the photon distributions will be presented below.

Finally the fringe SNR's at D1 are

$$\text{SNR}_{N_6}^+ = \sqrt{\frac{(m+(1+m)\,2\bar{n}\cos^2(\phi/2))^2}{(1+m)(1+2\bar{n}\cos^2(\phi/2))\,2\bar{n}\cos^2(\phi/2)}}, \quad m \geq 0$$

$$\text{SNR}_{N_6}^- = \sqrt{\frac{(1-m)\,2\bar{n}\cos^2(\phi/2)}{1+2\bar{n}\cos^2(\phi/2)}}, \quad m \leq 0$$

(4)

In Fig. 2 the Mathematical Protocol (MP) $P_6^+$ probability distributions and SNR are plotted for $m = 0, 1, 5, 10$ and $\bar{n} = 0.5$.

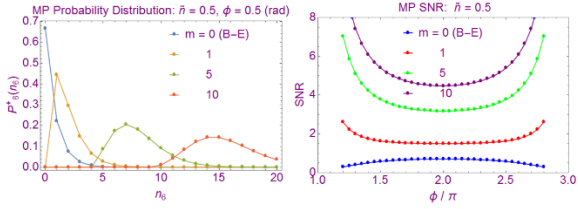

FIG. 2 MP Probability distributions and SNR at detector D1, see Fig. 3, for photon addition

The $m = 0$ B-E distribution is monotonically decreasing. For $m \geq 1$ (photon addition) the distribution acquires a bell-shaped form.

From the figure it is also obvious that the average occupation number increases when $m > 0$. It does so more rapidly with addition rather than subtraction (not shown) of photons. The MP photon count SNR grows significantly with increasing m values. The benefits (not shown) are however weaker in the case of photon subtraction.

The MP results have been applied to the simulation of imaging of a galaxy. The image reconstruction, as seen in Fig. 3, is based on Eq. (1) where the SNR is given in Eq. (4). The results indicate a clear improvement from $m = 0$ to $m = 10$. Several features of the galaxy that are unresolved at $m = 0$, become much more discernable at $m = 10$.

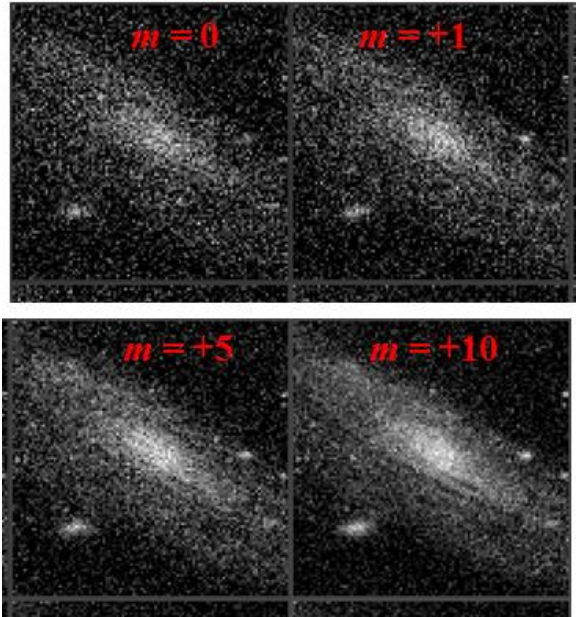

FIG. 3. MP numerically reconstructed image of a galaxy with $\bar{n} = 0.5$ and m=0, 1, 5, 10.

The MP used here is a mathematical abstraction. We need to devise a physical protocol (PP) that performs either photon subtraction or addition. The simplest PP is based on photon subtraction via a beam splitter and post-selection. We refer to it as PP-BS. In Fig. 4 we show the PP-BS schematic.

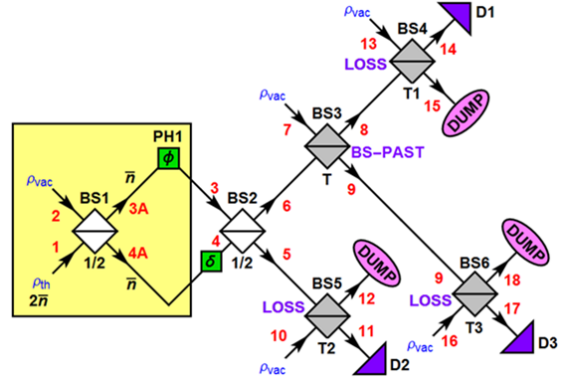

FIG. 4 Schematic for the simulation of PP-BS. BS4, BS5, BS6 are the beam splitters simulating the detector losses.

BS3 occasionally performs photon subtraction, and the D3 detector performs post-selection. We included losses, via beam splitters BS4, BS5, BS6 with transmission $T_1$, $T_2$ and $T_3$ respectively, to accurately emulate the experimental results. We analyzed two cases: a) post-selection measurement in which $m$ photons are subtracted; b) "click" detection, where any number, larger than zero, of photons are subtracted. We use the same MP methodology to study the PS-BS case, and briefly summarize the results below. The probability of success for m subtracted photons and the photon probability distribution are, respectively

$$p_m(T) = \frac{\left(2(1-T)T_3\,\bar{n}\cos^2(\phi/2)\right)^{-m}}{\left(1+2(1-T)T_3\bar{n}\cos^2(\phi/2)\right)^{1-m}}, \quad m < 0$$

$$P_{14}^-(n) = \frac{2^n}{(-m)!\,n!}\left(\bar{n}TT_1\cos(\phi/2)^2\right)^n\left(1+2\bar{n}(1-T)T_3\cos(\phi/2)^2\right)^{1-m}$$
$$\times\left(1+2\bar{n}(T(T_1-T_3)+T_3)\cos(\phi/2)^2\right)^{1+m-n}(-m+n)!$$

In the PP-BS with $m = -1$ case, the distribution shift is mitigated by the presence of losses. In the absence of losses PP-BS $\Rightarrow$ MP (for $m$=-1), hence the relevance of mitigating the losses. Finally the click detection distribution and SNR (not shown) are quite similar to the PP-BS for low $\bar{n}$.

The corresponding PP-BS SNR are

$$(SNR_{N_{14}})_0 = \sqrt{\frac{2\bar{n}TT_1\cos^2(\phi/2)}{1+2\bar{n}TT_1\cos^2(\phi/2)}}$$

$$(SNR_{N_{14}})_m = \sqrt{\frac{(1-m)2\bar{n}TT_1\cos^2(\phi/2)}{1+2\bar{n}(TT_1+(1-T)T_3)\cos^2(\phi/2)}}, \quad m < 0$$

(4)

There is a tradeoff between SNR and the probability of success, where the larger the T the larger the SNR at the expense of a very low probability of success. In the experiments we choose T ≈ 0.8. This rules out the use of PP-BS for practical applications; it is, however, of great relevance to experimentally verify that, if photon subtraction could be performed in a deterministic process, it does improve the SNR of a thermal beam.

A corroborating experimental effort was performed at the University of Rochester and a succinct summary is given here. We use a CW laser whose light is passed through a rotating ground glass to create thermal statistics, coupled into a single-mode fiber (SMF) and then into an MZI. One of the output ports of the MZI is coupled to a multimode fiber (MMF) and onto an avalanche photo diode (APD) detector. The other output port leads to a half-wave plate (HWP) and polarizing beam splitter (PBS) to implement photon subtraction before terminating in two APDs. The full detailed description and analysis of the experiment can be found in[7].

We determine the photon statistics from the APD measurements. The thermal source has a coherence time of roughly one microsecond (adjusted to match that of the laser) whereas the APD dead time is significantly shorter at about 50 ns. Thus, for a relatively small (< 4) number of photons in one source coherence time, the number of APD pulses approximates well the true number of photons in a coherence time of the laser. As described above, the detector quantum efficiency can be taken account of in the modeling by adding another beam splitter to the path.

We generated the MZI fringes and evaluated the SNR. The results are shown in Fig. 5 which shows SNR as a function of MZI phase $\phi$. These results agree with the predictions of Eq. (5). The agreement

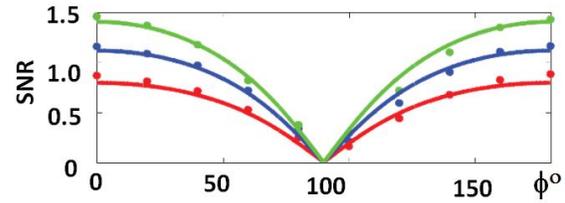

*FIG. 5 Comparison of theoretical predictions (lines) and experimental results (dots) for SNR as function of $\phi$ in degrees. Red, no photon subtraction, blue and greed one and two photons subtracted. With permission from [7]*

between theory and experiment is good and fully validates our conjecture that photon subtraction and, by induction, photon addition, lead to significantly improved SNR for single mode thermal beams.

We are working on a deterministic PP based on the injection of excited atoms in the thermal photons beam. A similar approach for photon subtraction and possibly photon addition is discussed in[8]. Here a photon is deterministically subtracted from a photon beam by a nearby atom due to destructive interference resulting in a Raman transfer of the atom from the ground state.

**Acknowledgements:** This research was developed with funding from the Defense Advanced Research Project Agency (DARPA) and by Boeing IR&D. The views, opinions, and/or findings expressed are those of the authors and should not be interpreted as representing the official views or policies of the Department of Defense or the U.S. Government. Approved for Public Release, Distribution Unlimited